# Half-Empty Offices in Flexible Work Arrangements: Why are Employees Not Returning?

Darja Smite[2,1], Nils Brede Moe [1,2], Anastasiia Tkalich[1] , Geir Kjetil Hanssen[1], Kristina Nydal[1], Jenny Nøkleberg Sandbæk[1], Hedda Wasskog Aamo[1], Ada Olsdatter Hagaseth[1], Scott Aleksander Bekke[1], and Malin Holte[1]

[1] SINTEF Digital, Trondheim, Norway
[2] Blekinge Institute of Technology, Karlskrona, Sweden

**Abstract.** Although the pandemic times of the world-wide forced working from home seem to be in the past, many knowledge workers choose to continue working predominantly from home as a partial or permanent practice. Related studies show that employees of companies from various industries, diverse in size and location, prefer to alter working in the office with working at home, coined as hybrid or flexible working arrangements. As a result, the post-pandemic times are associated with empty offices, confused managers and organizational leaders not knowing what to do with the often-expensive rental contracts. In this paper, we investigate the employee presence in the offices in two software companies and dive deeper into the reasons behind the preferences to work remotely, practices that help to attract employees back into the offices and, in cases when this is not possible, the ways companies can repurpose the office space for the future needs of their employees. The latter are based on the qualitative analysis of interviews and survey responses. Our findings suggest that since the fall 2021 the offices were half-empty and that, on average, the daily office presence varies between 15-30%. The reasons for working remotely include behavioural and practical motivations, as well as factors related to office equipment and facilities, and the nature of the work tasks. Finally, we discuss the practical implications of our findings on the future work arrangements.

## 1 Introduction

The forced work from home (WFH) during the pandemic in many software companies demonstrated that perceived productivity not only remains stable, but in some cases improves [1]. As a result of the more positive than expected experiences with remote working, many employees choose to continue working from home if not full time, than at least part-time, altering the days of office presence with days working from home [2], which is often referred to as *hybrid model* or *flexible work arrangements* [3]. As the offices remain half-empty (or hardly occupied) even after the reopening of the societies, there is a growing realization that flexible work arrangements are here to stay, marked as a work-life shift [3] or a revolution of telework [4].

While there are many benefits of working from home [2, 5, 6], remote communication weakens the connection between colleagues [7, 8] and makes non-verbal signals harder to notice, even in video meetings [9]. This significantly complicates the team managers' job, which often depends on the ability to observe, and communicate in-the-moment feedback with team members, engage in conversations and debate [3]. Evidently, this is not only about the Tayloristic managers who rely on the management style of command and control, but equally about the Participative and Supportive leadership styles prevalent in agile environments. Besides, continual face-to-face interactions are said to be important for commitment, assistance and collaboration, knowledge sharing [9]. Similarly innovation is dependent on face-to-face contact with customers and colleagues, who may generate ideas in planned and spontaneous brainstorming sessions and conversations [3]. These types of cooperation often find place in the hallways, by the coffee machines, water coolers, copiers or between meetings [3, 9]. It is evident that the question of attracting the employees back into the offices is of crucial importance, but how to achieve this in the times when even onboarding is performed remotely and the degree of flexibility offered by the company becomes the make-or-break point for many job seekers? What are the reasons for the shift from predominantly onsite work mode to close to entirely remote working? Motivated by the hybrid future challenges, in this paper, we seek to answer the following research questions:

**RQ1:** *How often are employees present in the office?*
**RQ2:** *What hinders and what motivates employees to visit the office?*

The rest of the paper is organized as follows. In Section 2 we outline the background and motivation for our study. Section 3 details the methodology and describes the empirical cases. In Section 4 we share our findings. Finally, Section 5 concludes the paper with responses to our research questions discussed in the light of related work, implications of our findings and future work directions.

## 2 Background

A growing number of organizations are implementing the flexible work arrangements for their employees, including Google, Telenor, Microsoft, and Spotify. Further, a large amount of research studying WFH have concluded that remote work *per se* does not hinder software engineers [10] and thus argues that hybrid work is here to stay, and will likely be the default work arrangement for software engineering in the future [1, 4, 6, 11]. However, little literature has attempted to predict how flexible arrangements should be performed in practice.

Researchers have reported more effective individual task solving and work coordination from the home office. Reasons include better focus time, fewer interruptions, more time to complete work, more efficient meetings, and a better/more comfortable work environment [5, 12]. Smite et al. [1] found fewer distractions and interruptions, increased flexibility to organize ones work hours, and easier access to developers a person depends on to complete the work. Ford et al. [5] report that not having to spend time commuting and having a more flexible schedule contributed to a better work-life balance.

On the other hand, there are clear weaknesses of working in isolation from the office and the colleagues. While individual tasks are solved more effectively, tasks that require coordination suffer and new tasks that require brainstorming are not easy to perform virtually [1]. Santos and Ralph [6] studied the implications of remote coordination when co-located teams work exclusively from home. They found that the needs for coordination increase when working remotely since group cohesion and communication were impaired by working from home. Santos and Ralph [6] worry that these challenges will persist in hybrid work and that hybrid teams may undermine agile processes. Smite et al. [1] identified weakened socialization and informal communication, team cohesion, problem-solving, and knowledge sharing as challenges posed by WFH. They argue that ”office-home” mixes are likely to increase these challenges. It is further predicted that information will likely circulate in the office without sufficiently reaching those who work from home [1].

In contrast with the short-term gains of productivity during the pandemic, companies are concerned about the alienation of colleagues and weakening of the knowledge networks as the employees continue working remotely [7]. Thus, some companies are introducing work policies that restrict the flexibility and constrain the number of days for working remotely, or introduce mandatory office days [13]. However, in practice, forced office presence can backfire with increased attrition levels. This is evident in a study that shows that 40% of employees who currently work from home, even if only one day a week, would seek another job if employers require a full return to the office [2]. Thus there is a growing interest in research that would shed light on the reasons why employees prefer to work remotely and what can change their motivation to return to the office.

## 3 Methodology

This paper presents a multiple-case holistic study [14], in which we study one phenomenon in two companies. The goal of this research is to understand whether we shall expect employees of software companies to return to the traditional work in the office, how occupied are the offices after the post-pandemic reopening of the society in Norway and what are the factors that influence the individual choices of the employees to work in the office or continue working remotely. Therefore, we collected data from two companies, Knowit and Sparebank 1 Dev, developing software-intensive products, sending their employees to work from home during the pandemic, and reopening their offices in fall 2021 with an episodic WFH advice during winter 2022. The choice of the companies was driven by convenience sampling, i.e., both companies are a part of an ongoing research project and had readily available data that helps to answer our research questions. Further, both companies are known as mature agile companies that are leading in experimenting with new practices and methods. As we both know the companies from before and they have mature development processes, we reduce the possibility of internal challenges and problems affecting developers' preferences for office use. Besides, both companies reported concerns regarding the return of the employees to the office work (case selection predicts similar results [ref to Yin]), which made them excellent candidates for our study with the opportunity for complementary findings that work together to enrich our understanding of the implications of the flexible work arrangements on the office presence. In the following, we outline the study context, and explain our data collection and analysis procedures.

### 3.1 Study context

**Knowit** is a large IT consultancy company with a large presence in the Nordic countries, but also other parts of Europe. The focus of this study was Knowit Objectnet, a subsidiary with approximately 175 consultants, where the main office is located in Oslo. In April 2022 they moved to brand new offices in downtown Oslo, the capital of Norway. The new office no longer offers a free parking for the employees. Consultants are mostly working for clients, often at the client site, but they may also work from home, or from the main Oslo office which was completely re-built and opened during the spring of 2022. As an HR manager from Knowit explains *"We have told our employees that they can decide completely themselves"*. This office deliberately has fewer work places than employees and is designed and equipped to be an arena for physical meetings, both for work but also very much for socializing. But this does not mean that the management prefers the remote working, as emphasized by the HR manager: *"[Our employees] work in teams so when they consider whether to sit at home or in the office, they have to prioritize the team's efficiency over their own efficiency. And then we [the management] are clear that we want the employees to be in the office and contribute to the environment. In other words, we are concerned with the culture, and it is more difficult to build a community when everyone is at home."*

**Sparebank 1 Dev** is a Fintech company developing software for Norwegian Banks. The organization offers a wide area of services and caters to both the consumer and professionals. Counting both their in-house employees and consultants, Sparebank 1 had 650 employees at the moment of our study. The teams had considerable freedom in how they worked. Most teams used a Kanban variant with Scrum elements and several deliberation practices, such as backlog meetings, team meetings, and daily standups. They used objectives and key results (OKRs) to guide their work and, influenced by Wodtke (2017), used "Monday commitments" and "Friday wins". Teams used retrospectives to improve work practices, and structured problem solving for continuous improvement. Finally, Sparebank 1 Dev regularly performed company surveys to understand the work-from-home situation. The bank renovated their office space during the pandemic.

### 3.2 Data collection

We have collected office presence data from the records of desk booking and access cards entries from two companies. Office presence data was extracted and analyzed in Excel quantitatively to visualize the office presence during the different weeks and months (Fig. 1 and Fig. 2) and to derive the average presence in the different weekdays (Fig. 3). The available data was quite simple, in the summary format and was not traceable to individuals. Hence, our analysis was limited to descriptive statistics.

Qualitative data regarding personal preferences for working in the office or working remotely have been obtained from qualitative interviews in Knowit and the employee surveys in Storebank 1 Dev.

In Knowit, we performed 12 semi-structured interviews (see the detailed overview of the interviewees in Table 2), which were transcribed and analysed using thematic analysis. The interview guide was based on the results of a literature summary, and a pilot interview was conducted. The researchers team printed all transcribed interviews and identified information or statements of interest, cut out paper snippets, and grouped these into themes that were given a descriptive name, e.g. "special reasons to be present at the office", or "improvement suggestions". A set of hypotheses emerged on the side as these groups were created, hence being a form of constant comparison [15]. This analysis resulted in a collection of 32 themes. As Knowit had moved into brand new offices two months before the interviews (see an overview of the offices in Fig. 4), the interviewees could reflect on their experience with both offices. The new offices had more seats and several social zones.

In Sparebank 1 Dev, we collected survey responses regarding the personal experiences with the chosen work arrangements. The survey received 244 responses from the maximum of 650 employees resulting in the response rate of 36%. The responses included the reasons for working remotely as opposed to not working in the office, which were extracted and comparatively analyzed together with the responses from the qualitative interviews conducted in Knowit.

An overview of data collection activities and data sources is available in Table 1.

**Table 1.** Overview of the data sources.

| Company | Data collection | |
|---|---|---|
| | **Office presence** | **Reasons for remote working** |
| Knowit | Archival data from a desk booking system Seatit during 2021-10-13 – 06-04-2022. | 12 semi-structured interviews about remote work preferences and office presence in June 2022. |
| Sparebank 1 Dev | Access card data during 2021-10-01 – 30-04-2022. | 244 survey responses about remote work preferences and office presence October and December 2021 |

**Table 2.** Overview of the interviewees at Knowit

| No | Role | Gender | Age group |
|---|---|---|---|
| 1 | Consultant and developer | Female | 35-40 |
| 2 | Director of business development | Male | 50-55 |
| 3 | Operation manager | Male | 45-50 |
| 4 | Designer | Female | 25-30 |
| 5 | HR manager | Female | 45-50 |
| 6 | Unit manager, sales, and business development | Male | 60+ |
| 7 | Director of product development | Male | 35-40 |
| 8 | Summer intern/part time consultant | Male | 25-30 |
| 9 | System developer | Male | 30-35 |
| 10 | System consultant | Male | 25-30 |
| 11 | Team lead | Male | 60+ |
| 12 | Frontend developer, senior developer | Male | 50-55 |

# 4 Results

In this section, we provide our findings with respect to office presence in the flexible work arrangements by starting with an overview of the office presence records (desk bookings and access card entries), followed by the analysis of the average presence during the different workdays and reasons for working from home vs working from the office.

## 4.1 Office presence

Our analysis of the office presence in both companies shows that the offices during the studied time period varied but has been relatively low (see Fig. 1 and Fig. 2). In the fall 2021 the office presence in Knowit was below 30% and below 50% in Sparebank 1 Dev. One exception in Sparebank 1 Dev was the third week of October when they organized an after-work social event with food, drinks and activities.

In winter 2021/2022, the new wave of the pandemic started and the employees were advised to work from home, which is evident in the low office presence during January 2022.

In spring 2022, the office presence in Knowit returned to the level of the fall 2021 in February until mid March and started to spike to close to 100% on certain days further on. This was motivated by the company-organized office-based social events. In Sparebank 1 Dev, both January and February were characterized by low office presence, which returned in the beginning of March to the level evidenced in the fall 2021 (around 50%).

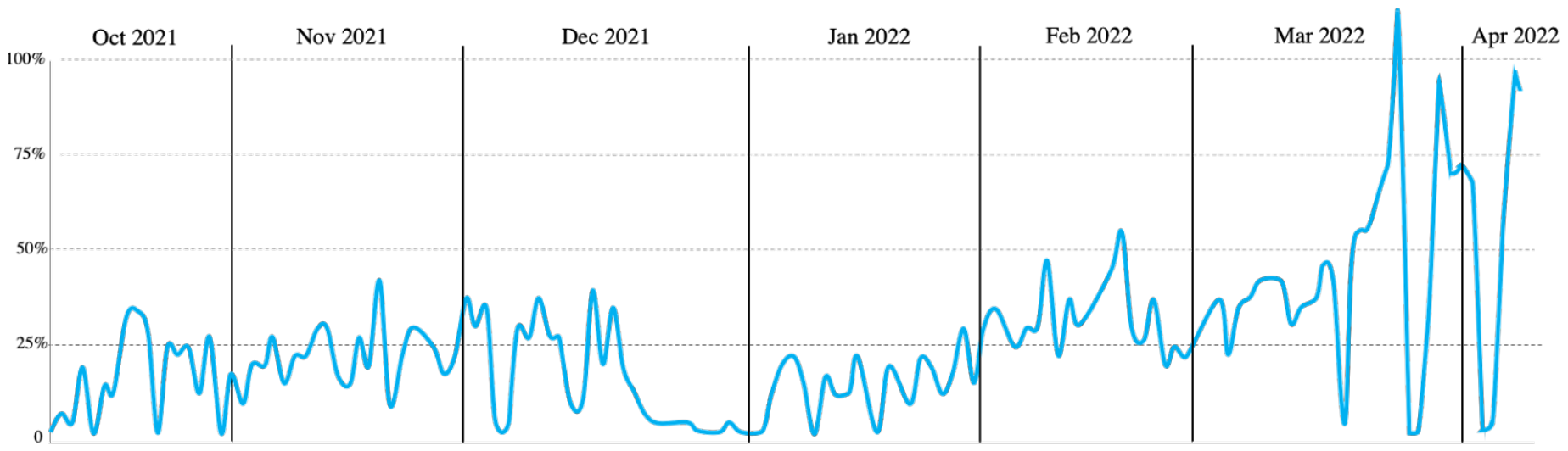

**Fig. 1.** Overview of office presence (desks booked per day, excludes days with no presence) in Knowit (Oct 2021 - Apr 2022).

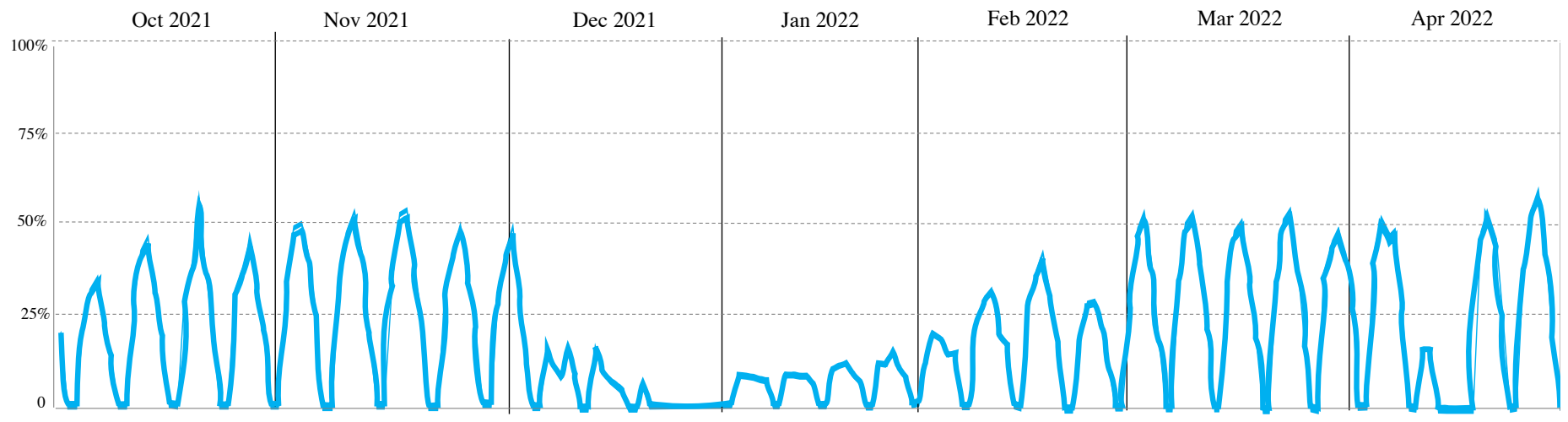

**Fig. 2.** Overview of office presence (access entries per day, including weekends and holidays) in SpareBank1 (Oct 2021 - Apr 2022).

When plotting the average number of employees in the office during the different workdays, we found different patterns across the two companies. In Knowit, the week started with the lowest office presence on Monday with the most popular office days being Wednesday, followed by Friday, when the company organizes afterwork activities. In contrast, the office presence in SpareBank1 resembled an even distribution curve with the highest presence on Wednesdays (typical days for social events) and lowest presence on Fridays.

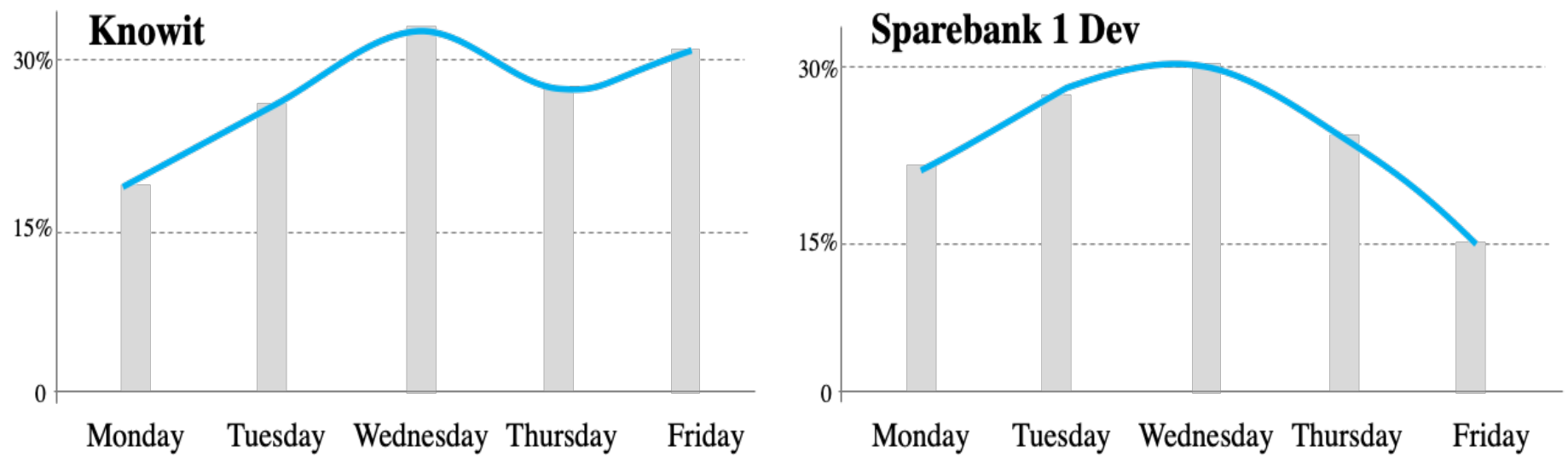

**Fig. 3.** Average office presence during the week.

### 4.2 Factors that motivate remote working

**Long or inconvenient commute to the office:** Long commute time was the number one reason for working remotely in Sparebank1 Dev, reported by more than 50% of the respondents. 30% commuted to work by train and 30% by tram or bus. Further 67% of respondents required more than half an hour to get to work (one way). The interviews in Knowit confirmed that the commute to work is one of the most important factors that motivate remote working. Knowit has recently moved their offices to a new building in the capital of Norway, Oslo, which does not provide any parking spaces. However, despite the convenient location, this meant that the employees had to walk, bike, or use public transportation to reach the offices, which was seen as more difficult than jumping into the car in the morning. Commute time also made a difference. In fact, two interviewees admitted that if they lived closer to the office, they would be more likely to work onsite. One experienced developer explained: "*I use a bus, an express bus. It takes about 45-50 minutes to work and over an hour back. It affects my preferences for working from home.*"

**Superior ability to focus at home:** Home office environment was reported to be superior in providing the ability to concentrate as reported by the survey responses from Sparebank1 Dev, as a respondent commented: *"For me, the noise [in the office] is a real problem. I use a lot of energy just to keep my concentration"*.

**Better-equipped home-office:** In addition to the already mentioned factors, investments into the home-office during the pandemic is what makes people comfortable to work from home and, might swing the weights when deciding whether to commute to the office. At least Knowit employees do not have to come to the office when occupied with individual tasks. Desks, chairs, and monitors that were acquired (and paid for by Knowit) during the pandemic all were kept by the employees. As one interviewee described, *"They [the company management] make sure that we have really good equipment. The most crucial things, such as furniture, desks, monitors, and more, they are really, really good"*. Similarly, several people from Sparebank 1 Dev have mentioned the problems with the equipment in the office and more superior conditions at home. Also Sparebank 1 Dev provided financial support for buying office equipment.

**Convenience for running personal routines at home:** Some interviewees mentioned various ways their daily or weekly needs are difficult to meet when being in the office. For example, for one interviewee it was about exercising. If there was a gym in the office, it would be more appealing to come into the office more often. Convenience for completing personal issues when working from home was also mentioned in Sparebank 1 Dev's survey.

**Habit of working from home:** Besides, the very habit of working from home during the pandemic times resulted in many continuing to work remotely, as reported by a large number of respondents in the Sparebank1 Dev's survey, as a respondent commented: *"I believe that people have become comfortable with home offices and the extra time it brings to the family etc."*.

**Schedule full of online meetings:** We found that some respondents in Sparebank1 Dev found it impractical to commute to the office when their workdays are packed with online meetings. As a respondent commented, *"[We have] challenges with equipment in meeting rooms – not adapted for having [hybrid] meetings in Teams. When you are* *a presenter, you need two screens. So when there are many Teams meetings it is more practical to sit at home."*

**Other reasons.** Among the other reasons that were mentioned by few of the research participants were better coffee machines, food, lighting and air ventilation in the household, and the good weather conditions. As one of the interviewees described: *"When the weather is nice, I rather stay at home"*.

### 4.3 Factors that motivate office presence

**Social interaction with colleagues:** Many interviewees revealed that the prime purpose for coming into the office was to meet colleagues and especially to get to know people, i.e., grow personal contract networks. This was also echoed in the survey responses from Sparebank 1 Dev. To make the offices more attractive both companies invested in upgrading them during the pandemic with the goal to create more social zones (see Fig. 4). Further, a few interviewees said that being in the office provided the unique feeling of inspiration from interactions with colleagues that was absent when working from home. One developer working at Knowit for two years explained *"I love being in the office. I enjoy working with people. I like meeting other people. I absolutely depend on that to do a good job, I depend on being happy, and then I want to meet other people, so I want to be in the office as much as possible."* During the pandemic people that strongly wanted to work from the office were allowed to do so. Especially those people who did not have access to any meeting places when the society was shut down, appreciated the possibility to meet colleagues in the office.

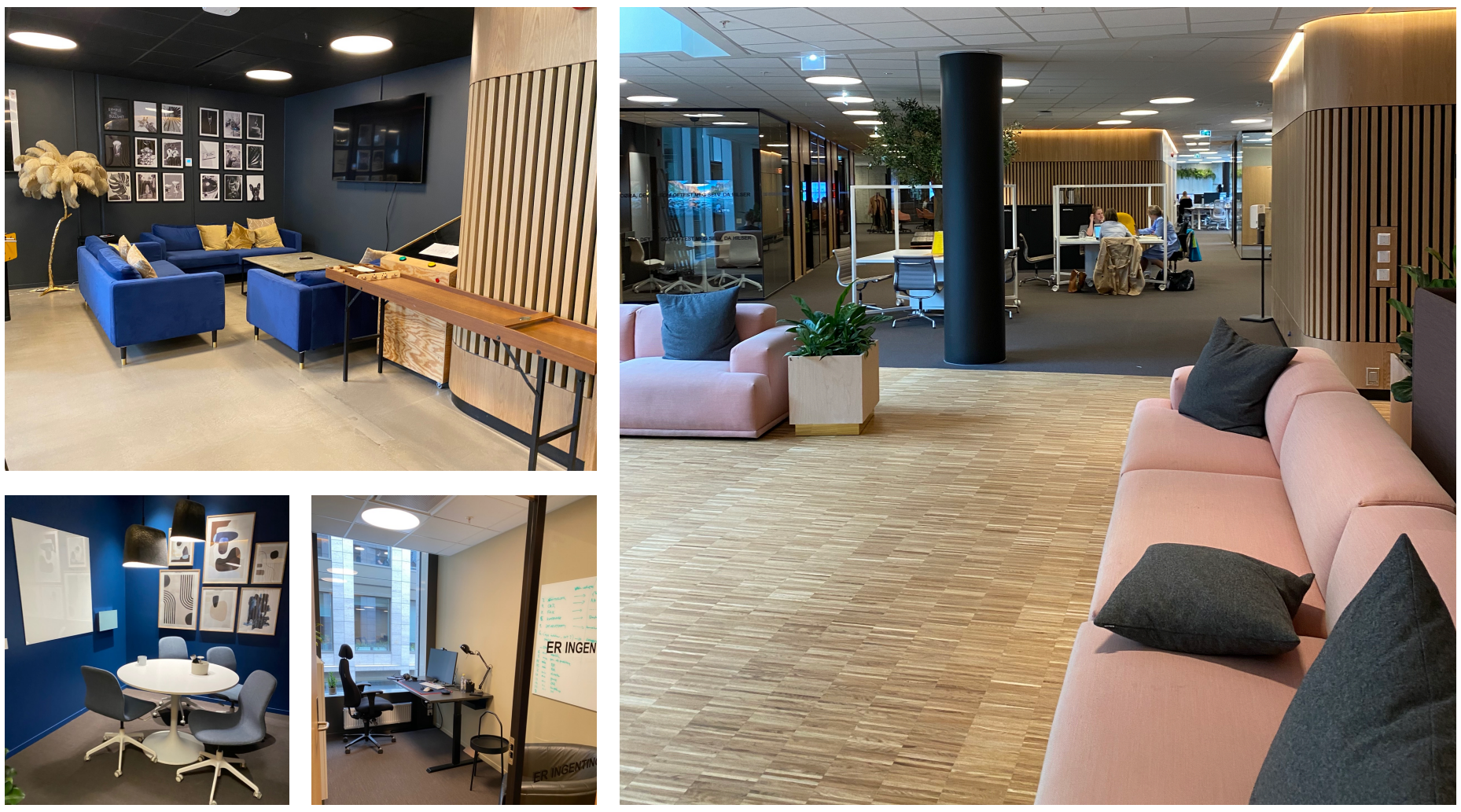

**Fig. 4.** The new offices of Knowit (April 2022) with several social zones, bookable meeting rooms, individual offices, and team areas in the open work area.

**Presence of colleagues in the office and their availability:** Many interviewees commented on the use of the desk booking system that allowed checking who else is planning to be in the office on a particular day. Colleagues' presence was reported to have a profound influence on the personal choices in the way that people were more likely to come in when colleagues of interest reported to be onsite and when their calendars showed that their schedules are not fully booked with meetings. A few interviewees confessed that they were also more likely to work from home if many other colleagues were not reporting to come into the office. This was also the case among a group of survey respondents in Sparebank1 Dev. Since the main motivation to be in the office is to socialise with others, nobody really wants to commute to the office to work in isolation. As an interviewee from Knowit explains: "*It's not only about knowing that Erik will be at work, but also that Erik actually has time for visitors*". A product manager further explains "*If I see that my colleague booked a desk, I book one next to him".*

**Tasks that require interaction:** Several interviewees mentioned that the nature of work and the nature of daily tasks might have a large influence on their decision to come to the office. For example, some meetings were easier to conduct and more productive when held in person, while individual tasks felt more appropriate for focused work in isolation at home. As one interviewee explained: "*So I'm trying to set up this work-from-home day for focused work, I need concentration then. The meetings are also best held in the office".* These meetings included planning meetings, brainstorming meetings, first customer meetings, workshops, and task assignment meetings.

**Enhanced work/life balance**: A few interviewees mentioned that they needed to go into the office to differentiate between work and home, in other words, improve the work/life balance. An interviewee who preferred working from the office explained: "*It's a lot easier for me to separate work from not working when I'm in the office, especially when it comes to breaks. When I'm at home, I find it a bit difficult. When you are taking a break at home, do you work or do you not work? Yes, it gets a little weird as it blurs a little into each other".*

**Other reasons:** Finally, some of the interviewees mentioned reasons that would occasionally influence their choice for commuting to the office, such as the need to leave the house and additional errands planned that require commuting anyway, which all increased the likelihood of working in the office but were not permanent motivators. Besides, a few research participants speculated that better food in the canteens, coffee machines and availability of a gym in the office would make them change their mind and come more often to the office.

## 5 Concluding discussion

In this paper, we have explored the actual state of the office presence in two software companies operating in Norway and the reasons for continuing working remotely after the reopening of the society after the pandemic. Both companies are known for being mature when it comes to agile methods and techniques, they upgraded their office space during Covid-19. The current work policy in the companies was that developers could spend two days at home and three days at the office, and this could be adjusted after agreeing with the work unit or the team, the customer and the manager. Further, at Knowit, employees need to have a suitable home-office to be allowed to work from home.

### 5.1 Employee presence in the office

With respect to our first research question (RQ1: How often are employees present in the office?), we found that the offices have been half-empty. The office presence in both companies have been below 50% since the fall 2021 with the lowest attendance in the winter months, January and February 2022, when a wave of infections increased. The office attendance increased only in one of our cases, in Knowit, as a result of the realization of the long-term impacts of remote working and the importance of social interaction. Our findings provide field evidence that supports the results of surveying employee preferences for continuing to work remotely [2, 13].

However, it is also fair to state that we are still witnessing the transition from the forced WFH during the pandemic to what we believe will be the hybrid work arrangements of the future. As a respondent from Sparebank 1 Dev explains: *"It's difficult to go from 100% home office to 100% office in such a short time, now Norway just opened too, and before that there was a lot of infection, so a longer transition phase is needed".*

### 5.2 Factors that influence where people work from

With respect to our second research question (RQ2: What hinders and what motivates employees to visit the office?), we found a list of factors that motivate remote work and a list of factors that motivate office presence (see a summary in Table 3). Notably, some of the reasons are behavioural, some have practical motivation, some are related to the office equipment and design, and some are dictated by the nature of the work tasks. The main driver for working from home is the commute. One implication of this is that if a company wants a high office presence, they need to be located in a place that is easy and fast to reach. Further, the main driver for being at the office is meeting the colleagues in your team or your group, i.e., the people one works with. If one's colleagues are not there, it is more likely that one will stay at home. As we found that social activities right after work motivate people to come to the office, social arrangements can be used as a tool to increase office presence. While all developers and team members appreciate focused time alone, they appreciate being with their colleagues as well. Further, our findings indicate that employees might be more likely to work onsite in the offices that satisfy the majority of their needs, including zones for collaboration and social interaction and silent zones for focused, undisturbed work, good quality office equipment, and good quality food in the canteen.

Yet, our findings are likely to indicate that hybrid work arrangements, in which the office days are mixed with the WFH days, are likely to remain the trend for the future, since this is an easy way to satisfy the diverse needs of the employees. This is also consonant with prior studies that demonstrated that working from home has both its advantages and disadvantages [1, 3, 5, 7, 10, 12]. A fair implication of this is that companies might consider reducing or repurposing their office space, especially on the days the offices are mostly empty, since the office presence never reaches 100%. The solutions to this problem include hot desking, having onsite work divided into shifts, moving to smaller offices or maybe renting out company offices to startups or partners on the days with the lowest attendance.

**Table 3.** Factors motivating remote work and office presence.

| Factors motivating remote work | Knowit | Sparebank 1 Dev | Factors motivating office presence | Knowit | Sparebank 1 Dev |
|---|---|---|---|---|---|
| Long or inconvenient commute to the office | X | X | Social interaction with colleagues | X | X |
| Superior ability to focus at home (noise and interruptions in the office) | | | Presence of colleagues in the office and their availability | X | X |
| Better-equipped home office | X | X | Tasks that require interaction | X | |
| Convenience for running personal routines while at home | X | X | Enhanced work/life balance | X | |
| Habit of working from home | | X | Need to leave the house | X | |
| Schedule full of online meetings | | X | Additional errands planned that require commuting anyway | X | |
| Better conditions at home: coffee, food, lighting or ventilation | | X | | | |
| Good weather conditions | X | | | | |

## 5.3 Future work

Our exploratory findings show a need for a deeper understanding of what are the good team and company strategies when introducing flexible work arrangements. It is evident that when software developers, product owners, or managers know that their colleagues will be working onsite, it is more likely that they will show up themselves. Therefore, more research is needed to study desk booking systems and other systems providing visibility into the office presence. Further, the current task a team member is working on affects the preferences for where to work from. Our related study [16] confirms that developers chose to perform tasks with vague requirements in co-location while individual tasks that require focus are best performed at home. Future research shall explore how to plan and organize a hybrid work week optimized for individuals, the team, and the company. As long commute time is the main driver for working from home, more research is needed to understand how team members' geographical distance to the office affects hiring and team composition strategies. Should companies that expect high office presence employ people only living near the office? Should companies compose teams based on the member location and with similar office presence preferences in mind? Finally, future research shall also explore how to onboard new team members, given that traditionally new hires are onboarded through close onsite mentoring requiring high office presence both from the new people and their senior team members.

## Acknowledgements

We thank Knowit AS and Sparebank 1 Dev for their engagement in our research, and the Norwegian Research Council for funding the research through the projects Transformit (grant number 321477) and 10xTeams (grant number 309344).